\documentclass[prd,eqsecnum,superscriptaddress, showpacs,nofootinbib]{revtex4}

\global\arraycolsep=2pt
\usepackage{stmaryrd}
\usepackage{amsmath}
\usepackage{amssymb}
\usepackage{graphicx}
\usepackage{textcomp}
\usepackage{calrsfs}
\usepackage{yfonts}
\usepackage{bm}
\usepackage{color}

\newcommand{\ben}{\begin{equation*}}    
\newcommand{\een}{\end{equation*}}
\newcommand{\bean}{\begin{eqnarray*}}      
\newcommand{\eean}{\end{eqnarray*}}

\newcommand{\nn}{\nonumber}
\newcommand{\be}{\begin{equation}}  
\newcommand{\ee}{\end{equation}}
\newcommand{\bea}{\begin{eqnarray}} 
\newcommand{\eea}{\end{eqnarray}}
\DeclareMathOperator{\Tr}{Tr}
\DeclareMathOperator{\tr}{tr}
\begin{document}

\title{Negative Casimir Entropies in Nanoparticle Interactions}

\author{K. A. Milton}
\email{kmilton@ou.edu}
\affiliation{H. L. Dodge Department of Physics and Astronomy,
University of Oklahoma, Norman, OK 73019 USA}
\affiliation{Laboratoire Kastler-Brossel, CNRS, ENS, UPMC, Case 74,
F-75252 Paris, France}
\author{Romain Gu\'erout}
\email{romain.guerout@upmc.fr}
\affiliation{Laboratoire Kastler-Brossel, CNRS, ENS, UPMC, Case 74,
F-75252 Paris, France}
\author{Gert-Ludwig Ingold}
\email{gert.ingold@physik.uni-augsburg.de}
\affiliation{Institut f\"ur Physik, Universit\"at Augsburg, 
Universit\"atsstra\ss e 1, D-86135 Germany}
\author{Astrid Lambrecht}
\email{astrid.lambrecht@upmc.fr}
\affiliation{Laboratoire Kastler-Brossel, CNRS, ENS, UPMC, Case 74,
F-75252 Paris, France}
\author{Serge Reynaud}
\email{serge.reynaud@upmc.fr}
\affiliation{Laboratoire Kastler-Brossel, CNRS, ENS, UPMC, Case 74,
F-75252 Paris, France}

\date{\today}

\begin{abstract}
Negative entropy has been known in Casimir systems for some time.  For example,
it can occur between parallel metallic plates modeled by a realistic Drude
permittivity.  Less well known is that negative entropy can occur
purely geometrically, say between a perfectly conducting sphere and a
conducting plate. The latter effect is most pronounced in the dipole 
approximation, which is reliable when the size of the sphere is small compared
to the separation between the sphere and the plate.  Therefore,
here we examine cases where negative entropy can occur
between two electrically and magnetically polarizable nanoparticles or atoms, 
which need not
be isotropic, and between such a small object and a conducting plate.
Negative entropy can occur even between two perfectly conducting spheres,
between two electrically polarizable nanoparticles 
if there is sufficient anisotropy,
between a perfectly conducting sphere and a Drude sphere,
and between a sufficiently anisotropic  electrically polarizable nanoparticle
 and a transverse magnetic conducting plate.
\end{abstract}

\pacs{42.50.Lc, 32.10.Dk, 05.70.-a, 11.10.Wx}

\maketitle
\section{introduction}
For more than a decade there has been a controversy surrounding entropy in the
Casimir effect.  This is most famously centered around the issue of how to 
describe
a real metal, in particular, the permittivity at zero frequency.  The
latter determines the temperature corrections to the free energy, and 
hence the entropy.  The Drude model, 
and general thermodynamic and electrodynamic 
arguments, suggest that the transverse electric (TE) reflection coefficient
at zero frequency
for a good, but imperfect metal, should vanish, while an ideal metal, or
one described by the plasma model (which ignores dissipation) has this
zero frequency 
reflection coefficient equal unity.  Taken at face value, the first,
more realistic scenario,  means that the entropy would not vanish at zero 
temperature, in violation of the Nernst heat theorem,
and the third law of thermodynamics.  However, subsequent
careful calculations showed that at very low temperature the free energy 
vanishes quadratically in the temperature, thus forcing the entropy
to vanish at zero temperature.  However, there would persist a region at low
temperature in which the entropy would be negative.  This was not thought
to be a problem, since the Casimir free energy does does not describe the
entire system of the Casimir apparatus, whose total entropy must necessarily
be positive.  However, the physical basis for the negative entropy region
remains mysterious.  For discussions of these effects see 
Refs.~\cite{bostrom04,oai:arXiv.org:quant-ph/0605005,ellingsen07,
oai:arXiv.org:0710.4882,dedkov08,gialsr,bordag10},
and references therein.

More recently, negative entropy has been discovered in purely geometrical
settings \cite{oai:arXiv.org:0911.0913}.  
Thus, in considering the free energy between a perfectly conducting
plate and a perfectly conducting sphere, it was found that when the distance
between the plate and the sphere is sufficiently small, the room-temperature
entropy turns negative, and that the effect is enhanced for smaller spheres.
For a very small sphere, the free energy and entropy are well-matched by a 
dipole approximation \cite{cdmnlr,lopez}.

The previous discussion
 suggests that this phenomenon should be studied in a systematic way.
In this paper we consider the retarded
 Casimir-Polder interactions \cite{Casimir:1947hx}
between a small object, such as a nanosphere or nanoparticle,
possessing anisotropic electric and magnetic polarizabilities, and a
conducting plate, and we analyze the contributions to the free energy
and entropy for the TE and TM (transverse magnetic)
 polarizations of the conducting plate.  The case of a small perfectly
conducting sphere above a plate is recovered by setting the electric
polarizability, $\alpha$, 
equal to $a^3$, where $a$ is the radius of the sphere,
and the magnetic polarizability, $\beta$, equal to $-a^3/2$.  We also examine
the free energy and entropy between two such anisotropically polarizable 
nanoparticles.
We find negative entropy not only as an interplay between TE and TM
polarizations in the plate, but even between a purely electrically polarizable
nanoparticle and the TM polarization of the plate, provided the nanoparticle 
is sufficiently
anisotropic.  The previous negative entropy results are verified, and we
show that even between electrically polarizable nanoparticles, 
negative entropy occurs
when the product of the temperature with the separation is sufficiently
small, provided the nanoparticles are sufficiently anisotropic.
The interaction between two identical isotropic small spheres modeled as
perfect conductors gives a negative entropy region, but not when they
are described by the Drude model (no magnetic polarizability); 
but the interaction between an isotropic perfectly conducting sphere and
an isotropic  Drude sphere gives negative entropy.
For room temperature, the typical distance at which negative entropy occurs is
below a few microns.

Negative entropy between an electrically polarizable atom and a conducting
plate was discussed in the isotropic case several years ago \cite{bezerra},
and the extension to a isotropic magnetically polarizable atom was
sketched in Ref.~\cite{oai:arXiv.org:0904.0234}.  The effects of atomic
anisotropy and of the different polarizations of the conducting plate were not
considered there.  The zero-temperature Casimir-Polder 
interaction between atoms having both isotropic
electric and magnetic polarizabilities was studied by Feinberg and 
Sucher \cite{feinberg}, while the temperature dependence for isotropic
atoms interacting only through their electric polarizability was first obtained
by McLachlan \cite{lachlan,lachlan2}.  Barton performed the generalization for
the magnetic polarizability at finite temperature \cite{barton}.
Haakh et al.\ more recently discussed the magnetic Casimir-Polder interaction
for real atoms \cite{haakh}.
The anisotropic case at zero temperature for the electrical Casimir-Polder
interaction was first given by Craig and Power \cite{craig,craig2}.
Forces between compact objects, which could include nanoparticles in
the dipolar limit, have been considered by many authors, for example in
Refs.~\cite{emig06,emig07,sernelius08,sernelius08a}, 
but less attention has been given
to the equilibrium thermodynamics of such objects interacting.

In this paper we consider anisotropic small objects, 
with the symmetry axis of the
objects coinciding with the direction between them or the normal to the 
plate, with both electric and magnetic polarizability.  Because we are 
interested in matters of principle, we work in the static approximation,
so both polarizabilities are regarded as constant, whereas most real atoms
have very small, and complicated, magnetic polarizabilities.  We also are
not concerned here with the fact that achieving large anisotropies is 
likely to be difficult for real atoms \cite{rep3}, because it may be
much more feasible to achieve the necessary anisotropies with nanoparticles,
such as conducting  needles. 

 We will work entirely in the dipole approximation for the 
nanoparticles, which is sufficient for large enough distances;
for short distances higher multipoles become important 
\cite{noguez03,noguez04}. We also ignore any possibility of temperature 
dependence of the polarizabilities.

We use natural units $\hbar=c=k_B=1$, and Heaviside-Lorentz units for
electrical quantities, except that polarizabilites are expressed in 
conventional Gaussian units.  

\section{CP free energy between a nanoparticle and a conducting plate}

We start by considering
 an anisotropic electrically and magnetically polarizable
nanoparticle 
a distance $Z$ above a perfectly conducting plate. We can take as our
starting point the multiple scattering formula for the interaction 
free energy between two bodies \cite{yale}
%\footnote{The relationship between the
%free energy and the entropy is clarified in the Appendix.} 
%(self-energies subtracted) 
\be
F_{12}=\frac12 \Tr \ln(\bm{1}-\bm{\Gamma}_0\mathbf{T}_1^E\bm{\Gamma}_0
\mathbf{T}_2^E)+\frac12 \Tr \ln(\bm{1}-\bm{\Gamma}_0\mathbf{T}_1^M\bm{\Gamma}_0
\mathbf{T}_2^M)-\frac12\Tr\ln(\bm{1}+\bm{\Phi}_0\mathbf{T}^E\bm{\Phi}_0
\mathbf{T}^M),\label{ms}
\ee
where the $\bm{\Gamma}_0$ is the free electric Green's dyadic,
\be \bm{\Gamma}_0(\mathbf{r,r'})=(\bm{\nabla\nabla}-\bm{1}\nabla^2)G_0(
|\mathbf{r-r'}|),\quad G_0(R)=\frac{e^{-|\zeta|R}}{4\pi R},\label{fgd}
\ee
in terms of the imaginary frequency $\zeta$.  The auxiliary Green's dyadic
is
\be \bm{\Phi}_0=-\frac1\zeta \bm{\nabla\times\Gamma}_0.
\ee
$\mathbf{T}^{E,M}_{1,2}$ 
are the electric and magnetic scattering operators for the
two interacting bodies.  Unfortunately, the EM cross term 
[the third term in Eq.~(\ref{ms})] in general does not
factor into separate parts referring to each body; $\mathbf{T}^{E,M}$ refer
to the whole system.  The trace (denoted $\Tr$) 
 includes an integral (at zero temperature) or
a sum (for positive temperature) over frequencies, and an integral over spatial
 coordinates, as well as a sum over matrix indices.  When the sum over
only the latter is intended, we will denote that trace by $\tr$.

For the case of a tiny object, it suffices to use the single-scattering 
approximation, and replace the scattering operator by the potential
\be
\mathbf{T}^E_n=\mathbf{V}^E_n=4\pi \bm{\alpha}\delta
(\mathbf{r-R}),\quad
\mathbf{T}^M_n=\mathbf{V}^M_n=4\pi \bm{\beta}\delta
(\mathbf{r-R}),
\ee
for a nanoparticle
 at position $\mathbf{R}$ with electric (magnetic)  polarizability 
tensors $\bm{\alpha}$ ($\bm{\beta}$). 
The approximation being made here is that the
nanoparticle is a small object, and it is adequate to ignore higher multipoles.
That is justified if $a$, a characteristic size of the particle, 
is small compared with the separation, $a\ll Z$.
Therefore, since at least one of our bodies is a nanoparticle, 
it suffices to expand
the logarithms in Eq.~(\ref{ms}) and retain only the first term.
Then we are left with the following formula for the Casimir-Polder free energy
between a polarizable nanoparticle and a conducting plate,
\be
F_{np}=-2\pi \Tr \left(\bm{\alpha} \bm{\Gamma}_0\mathbf{T}_p\bm{\Gamma}_0
+\bm{\beta}\bm{\Phi}_0\mathbf{T}_p\bm{\Phi}_0\right).\label{fap}
\ee
Here $\mathbf{T}_p$ is the purely electric scattering operator for the
conducting plate, which is immediately written in terms of the Green's operator
$\bm{\Gamma}$ for a perfectly conducting plate,
\be \bm{\Gamma}_0\mathbf{T}_p\bm{\Gamma}_0=\bm{\Gamma-\Gamma}_0.
\ee

\subsection{$\bm{\alpha}$ polarization of nanoparticle}
It is well-known \cite{levine} that the Green's dyadic for a perfectly 
conducting plate lying in the $z=0$ plane is for $z>0$ given by the image
construction
\be
(\bm{\Gamma-\Gamma}_0)(\mathbf{r,r'})=-\bm{\Gamma}_0(\mathbf{r},\mathbf{r}'
-2\mathbf{\hat z}z')\cdot(\bm{1}-2\mathbf{\hat z\hat z}),
\ee
where the free Green's dyadic is given by Eq.~(\ref{fgd}). Explicitly, the
latter can be written as \cite{tomsk}
\be  
\bm{\Gamma}_0(\mathbf{r,r'})=-[\bm{1}u(|\zeta|R)-\mathbf{\hat R\hat R}
v(|\zeta| R)]\frac{e^{-|\zeta| R}}{4\pi R^3},\quad \mathbf{R=r-r'},\label{fgd2}
\ee
in terms of the polynomials
\be
u(x)=1+x+x^2,\quad v(x)=3+3x+x^2.\label{poly}
\ee
Let us first consider zero temperature.  Then, if we ignore the frequency
dependence of $\bm{\alpha}$,  we integrate over imaginary
frequency, and we immediately obtain the famous Casimir-Polder result
\cite{Casimir:1947hx}
\be
E^E_{np}=-\int_{-\infty}^\infty d\zeta \tr \bm{\alpha \cdot(\Gamma-\Gamma}_0)
(\mathbf{R,R})=-\frac{\tr \bm{\alpha}}{8\pi Z^4}.
\ee

\begin{figure}
\includegraphics{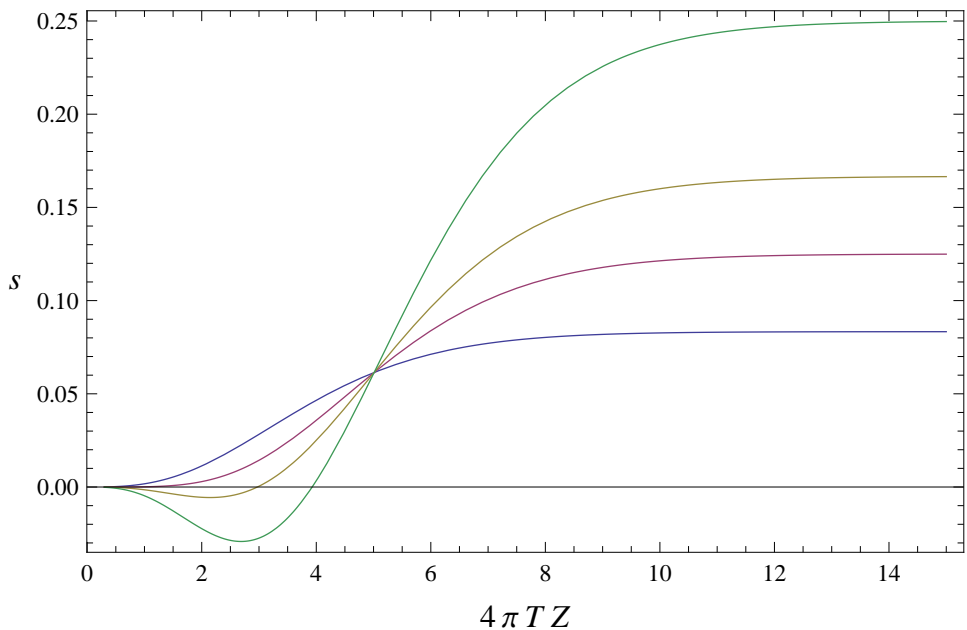}
\caption{\label{fig1} Scaled entropy $s$
between a purely electrically polarizable nanoparticle
and a conducting plate, as a function of the product of the temperature times
the distance from the plate.  The different curves (bottom to top for large
$ZT$) are for anisotropies $\gamma= 0$ (blue), $1/2$ (red), $1$ (yellow), 
$2$ (green). [Color online]}
\end{figure}

For nonzero $T$, we replace the integral by a sum,
\be
\int_{-\infty}^\infty \frac{d\zeta}{2\pi}\to T\sum_{m=-\infty}^\infty,
\ee and replace the frequency by the Matsubara frequency \cite{matsubara}
%(evidently known long before Matsubara)
\be
\zeta\to \zeta_m=2\pi m T.
\ee
 We assume the principal axis of the nanoparticle aligns 
with the direction normal to the plate,
\be
\bm{\alpha}=\mbox{diag}(\alpha_\perp,\alpha_\perp,\alpha_z),
\ee
and define the anisotropy $\gamma=\alpha_\perp/\alpha_z$.  $\gamma\gg1$
means that the nanoparticle is mostly polarizable in the direction 
parallel (transverse) to the plate, while $\gamma\ll1$ means the 
nanoparticle is mostly polarizable in the direction normal to the plate.
Then the free energy is easily obtained:
\be F^E_{np}=-\frac{3\alpha_z}{8\pi Z^4}f(\gamma,y),\quad
f(\gamma,y)=\frac{y}6[(1+\gamma)(1-y\partial_y)+\gamma y^2
\partial^2_y]\frac12\coth\frac{y}2 
\ee (the normalization is chosen so that $f(1,0)=1$), 
where $y=4\pi Z T$,  $Z$  being the distance between the
nanoparticle and the plate. The entropy is 
\be
S_{np}^E=-\frac{\partial}{\partial T}F_{np}^E
=\frac{3\alpha_z}{2 Z^3}\frac\partial{\partial y}f(\gamma,y),
\ee
so we define the scaled  entropy by
\be
s(\gamma,y)=\frac\partial{\partial y}f(\gamma,y).
\ee
For large $y$ this entropy approaches a constant, 
\be
s(\gamma,y)\sim \frac1{12}(1+\gamma),\quad y\to\infty,
\ee
while for small $y$,
\be
s(\gamma,y)\sim \frac1{540}(1-2\gamma)y^3+O(y^5).\label{ssmally}
\ee
  The entropy vanishes at $T=0$, and then starts off negative for small $y$ 
when
$\gamma>1/2$.  In particular, even for an isotropic, solely electrically 
polarizable, nanoparticle, where $\gamma=1$, the entropy is negative
for a certain region in $y$, as discovered in Ref.~\cite{bezerra}  
The behavior of the entropy with $\gamma$ is illustrated in  Fig.~\ref{fig1}.
For an isotropic nanoparticle, 
the negative entropy region occurs for $4\pi Z T<2.97169$,
or at temperature $300$ K, for distances less than 2 $\mu$m.

\begin{figure}
\includegraphics{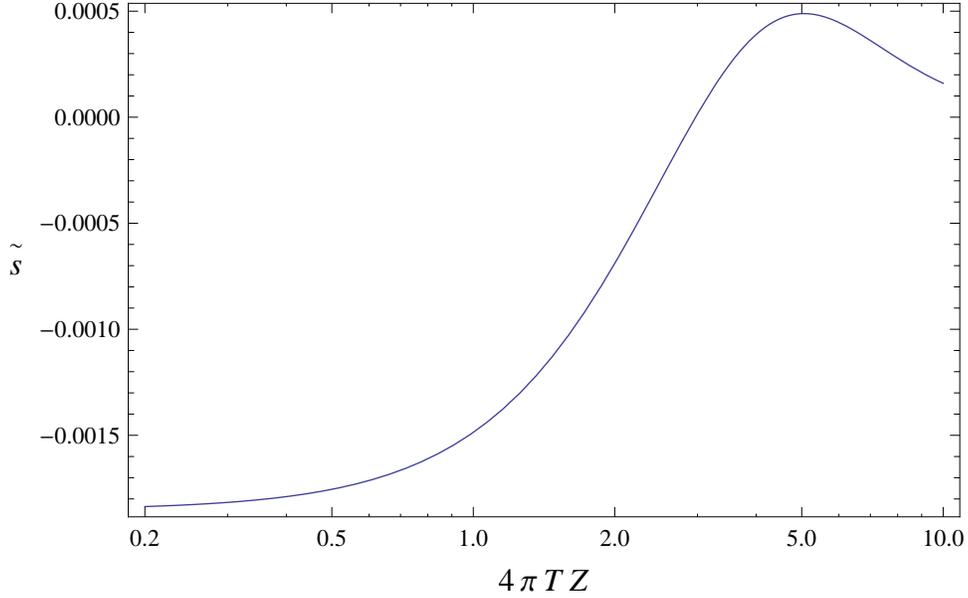}
\caption{\label{fig1a} Rescaled entropy $\tilde s$ for 
fixed temperature as a function of the distance of an isotropic
 atom from the plate.  The entropy tends to a finite negative
value for small distances, has a positive maximum, and then decreases to zero
from above for large distances.}
\end{figure}

Most Casimir experiments are performed at room temperature.  Therefore,
it might be better to present the entropy in the form
\be
S_{np}^E=\frac{3\alpha_z}2(4\pi T)^3\tilde{s}(\gamma,y),\quad
\tilde{s}(\gamma,y)=y^{-3} s(\gamma,y),
\ee
which in view of Eq.~(\ref{ssmally}) makes explicit that the entropy tends to
a finite value as $Z\to0$.  This version of the entropy for the
isotropic case is plotted in
Fig.~\ref{fig1a}.

\subsection{E and H  polarizations of plate}
\begin{figure}
\includegraphics{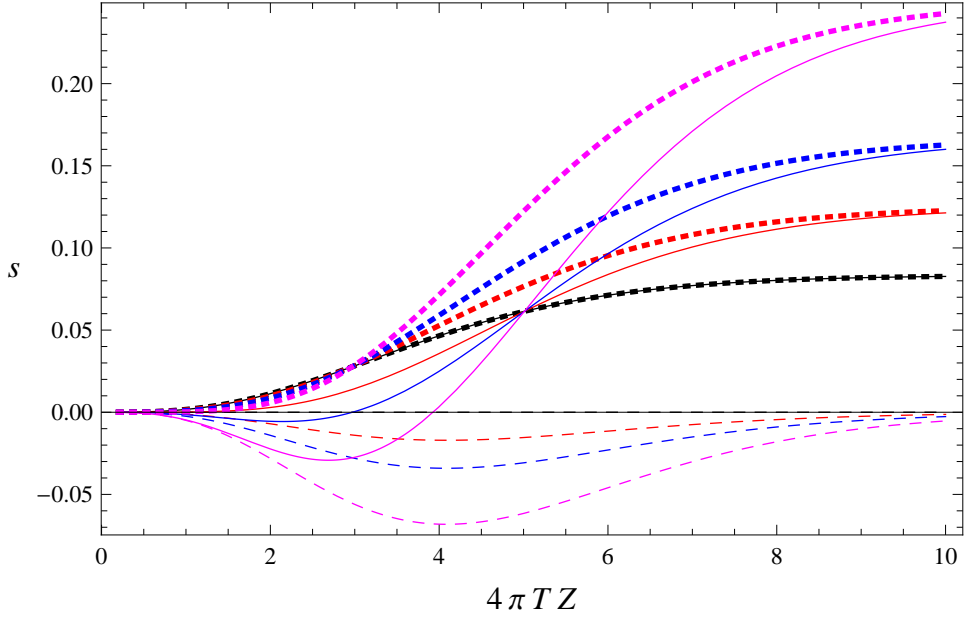}
\caption{\label{fig2} The entropy between an electrically polarizable 
nanoparticle
and a conducting wall.  The solid curves are the total entropy, the 
short-dashed curves are for the TM plate contribution, and the long-dashed 
curves are for TE.
Referring to the ordering for large $TZ$, the inner set of curves (black)
are for $\gamma=0$, the next set (red) is for $\gamma=1/2$, where the negative
total entropy region starts to appear, the third set (blue) is for $\gamma=1$,
and the outer set (magenta) is for $\gamma=2$. [Color online]}
\end{figure}

\begin{figure}
\includegraphics{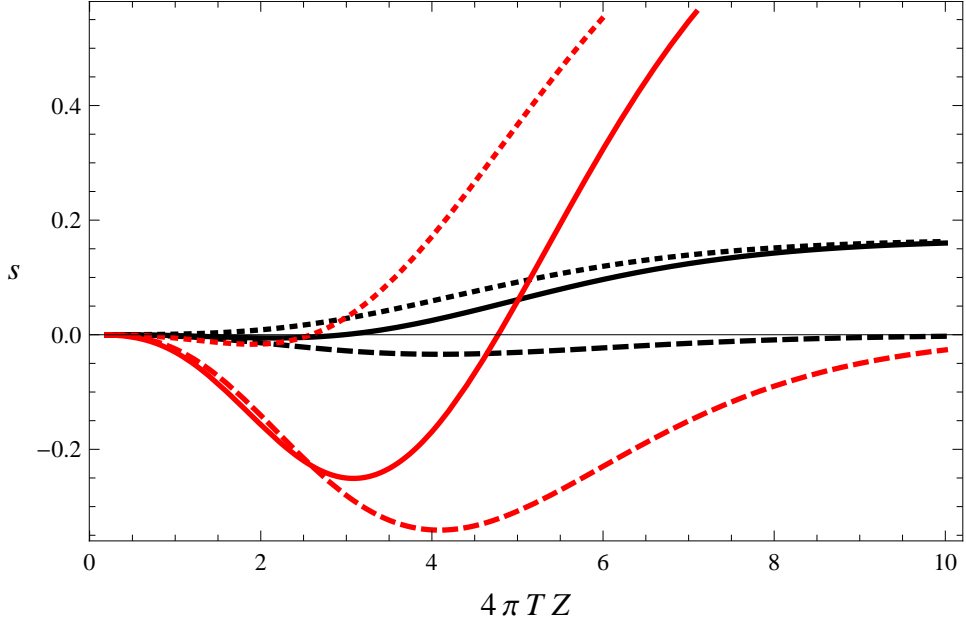}
\caption{\label{fig3} This illustrates that even for a solely electrically
polarizable nanoparticle $S_H$ can turn negative for $\gamma>2$.  The inner set
of curves (black) are for $\gamma=1$, and the outer curves 
(red) are for $\gamma=10$.
Again the total entropy is given by the solid curves, $s_E$ by the long-dashed
curves, and $s_H$ by the short-dashed curves. [Color online]}
\end{figure}

To understand this phenomenon better, let us break up the polarization states
of the conducting plate.  For this purpose, it is convenient to use the
$2+1$-dimensional breakup of the Green's dyadic.
Following the formalism in Ref.~\cite{rep3}, we find that
the free Green's dyadic has the form ($(d\mathbf{k}_\perp)=d^2k_\perp$)
\be
\bm{\Gamma}_0(\mathbf{r,r'})=\int\frac{(d\mathbf{k}_\perp)}{(2\pi)^2}
e^{i\mathbf{k_\perp\cdot(r-r')_\perp}}(\mathbf{\mathsf{E}+\mathsf{H}})(z,z')
\frac1{2\kappa}e^{-\kappa|z-z'|},
\ee
 which readily leads to 
the representation for the free energy for the nanoparticle-plate system
\be
F^E=2\pi T\sum_{m=-\infty}^\infty \int\frac{(d\mathbf{k_\perp})}{(2\pi)^2}
\tr [\bm{\alpha}\cdot
(\mathbf{\mathsf{E}-\mathsf{H}})(Z,Z)]\frac1{2\kappa}e^{-2\kappa Z},\label{fe}
\ee
where $\kappa^2=k_\perp^2+\zeta_m^2$. 
Here the TE and TM polarization tensors are, after averaging over the 
directions of $\mathbf{k_\perp}$,
\be
\mathbf{\mathsf{E}}
=-\frac{\zeta^2}2\bm{1}_\perp,\quad \mathbf{\mathsf{H}}=\frac{\kappa^2}2
\bm{1}_\perp+(\kappa^2-\zeta_m^2)\mathbf{\hat z\hat z}.
\ee
Performing the elementary integrals and sums, we get for the TE contribution
to the free energy
\be
F^E_E=-\frac{3\alpha_z}{8\pi Z^4}f_E(\gamma,y),
\quad f_E(\gamma,y)=\gamma\frac{y^3}{12}\partial_y^2 \left(
\frac12\coth\frac{y}2\right),
\ee
and to the entropy
\be
S^E_E=-\frac\partial{\partial T}F^E_E=\frac{3\alpha_z}{2Z^3}s_E(\gamma,y),
\quad s_E(\gamma,y)=\frac\partial{\partial y}f_E(\gamma,y).
\ee
For large $y$, $s_E$ goes to zero exponentially,
\be
s_E(\gamma,y)\sim-\frac\gamma{12}y^2(y-3)e^{-y},\quad y\gg1,
\ee
while for small $y$,
\be
s_E(\gamma,y)\sim -\gamma \frac{y^3}{360}+O(y^5), \quad y\ll 1.
\ee
The transverse electric contribution to the entropy, $s_E$, is always negative.
On the other hand, $s_H=s-s_E$ is positive for large $y$,
\be
s_H\sim \frac{1+\gamma}{12},\quad y\gg1,
\ee
but can change sign for small $y$,
\be
s_H(\gamma,y) \sim\frac{y^3}{540}\left(1-\frac12\gamma\right),\quad y\ll1.
\ee
So $s_H$ can change sign for $\gamma>2$; the total entropy $s$, in 
Eq.~(\ref{ssmally}), can change sign for
$\gamma>1/2$.  These features are illustrated in Figs.~\ref{fig2}, \ref{fig3}. 

Note that there is no difference between a perfectly conducting plate and
one represented  by the ideal Drude model, which differs from the former only
by the exclusion of the TE $m=0$ mode.  This is because this term does not
contribute to $F^E_E$ or to $S^E_E$.

\subsection{$\bm{\beta}$ polarization of nanoparticle}
Now we turn to the magnetic polarizability of the nanoparticle, 
that is, the evaluation of the
second term in Eq.~(\ref{fap}). Again, from Ref.~\cite{rep3}, all we
need is the scattering operator for the conducting plate,
\be
\mathbf{T}_p(\mathbf{r,r'})=\int\frac{(d\mathbf{k}_\perp)}{(2\pi)^2}e^{i\mathbf{k_\perp\cdot
(r-r')_\perp}}\frac1{\zeta^2}(\mathbf{\mathsf{E}-\mathsf{H}})(z,z')\delta(z)
e^{-\kappa|z'|}.
\ee
 Then the  Green's dyadic appearing there can be
written in terms of the polarization operators for the plate as
\bea
\bm{\Phi}_0\cdot\mathbf{T}_p\cdot\bm{\Phi}_0(Z,Z)&=&\int dz'\,dz'' \int
\frac{(d\mathbf{k_\perp})}{(2\pi)^2}\left(-\frac1\zeta\bm{\nabla}\times
(\mathbf{\mathsf{E}+\mathsf{H}})(Z,z')
\frac1{2\kappa}e^{-\kappa|Z-z'|}\right)\cdot\frac1{\zeta^2}
(\mathbf{\mathsf{E}-\mathsf{H}})(z',z'')\delta(z')\nn\\
&&\quad\cdot\left(-\frac1{\zeta^2}\bm{\nabla}''\times
\bm{\nabla}''\times-\bm{1}\right)e^{-\kappa|z''|}\cdot\left(-\frac1\zeta
\bm{\nabla}''\times (\mathbf{\mathsf{E}+\mathsf{H}})(z'',Z)
\frac1{2\kappa}e^{-\kappa|z''-Z|}\right).
\eea
The intermediate wave operator here annihilates the following Green's dyadic
except on the plate:
\bea
\left(-\frac1{\zeta^2}\bm{\nabla}''\times
\bm{\nabla}''\times-\bm{1}\right)e^{-\kappa|z''|}\cdot \bm{\nabla}''\times
(\mathbf{\mathsf{E}+\mathsf{H}})
&=&\left(\frac1{\zeta^2}\nabla^{\prime\prime2}-1\right)
e^{-\kappa|z''|}\bm{\nabla}''\times (\mathbf{\mathsf{E}+\mathsf{H}})\nn\\
&=&-\frac{2\kappa}{\zeta^2}\delta(z'')\bm{\nabla}''\times 
(\mathbf{\mathsf{E}+\mathsf{H}}).
\eea
Now we integrate by parts and use the identities \cite{Milton:2013xia}
\begin{subequations}
\bea
\bm{\nabla}'\times (\mathbf{\mathsf{E}-\mathsf{H}})(z',z'')
\times \bm{\nabla}''&=&
-\zeta^2(\mathbf{\mathsf{E}-\mathsf{H}}), \\
\mathbf{\mathsf{E}}(z,z')\cdot\mathbf{\mathsf{E}}(z',z'')
=-\zeta^2 \mathbf{\mathsf{E}}(z,z''),
\quad\mathbf{\mathsf{H}}(z,z')\cdot\mathbf{\mathsf{H}}(z',z'')
&=&-\zeta^2 \mathbf{\mathsf{H}}(z,z''),
\quad \mathbf{\mathsf{E}}(z,z')\cdot\mathbf{\mathsf{H}}(z',z'')=0.
\eea
\end{subequations}
In this way we find the magnetic Green's dyadic appearing in the
formula for the magnetic part of the Casimir-Polder energy (\ref{fap}) to be
\be
\bm{\Phi}_0\mathbf{T}_p\bm{\Phi}_0(Z,Z)=-\int\frac{(d\mathbf{k}_\perp)}{
(2\pi)^2}(\mathbf{\mathsf{E}-\mathsf{H}})(Z,Z)e^{-2\kappa Z},
\ee
which is just negative of the corresponding expression for the electric
Green's dyadic seen in Eq.~(\ref{fe}).  
Thus the expression for the magnetic polarizability 
contribution is obtained from the free energy for the electric polarizability
by the replacement $\bm{\alpha}\to -\bm{\beta}$, and the total free energy
for the nanoparticle-plate system is given by
\be
F=2\pi T\sum_{m=-\infty}^\infty \int\frac{(d\mathbf{k_\perp})}{(2\pi)^2}
\tr [(\bm{\alpha}-\bm{\beta})\cdot(\mathbf{\mathsf{E}-\mathsf{H}})(Z,Z)]
\frac1{2\kappa}e^{-2\kappa Z}.
\ee
This simple relation between the electric and magnetic polarizability
contributions was noted in Ref.~\cite{oai:arXiv.org:0904.0234}.
In particular, for the interesting case of a conducting sphere, the previous
results apply, except multiplied by a factor of 3/2.  In that case,
the limiting value of the entropy is 
\be
S(T)\sim -\frac4{15}(\pi a T)^3,\quad 4\pi Z T\ll1.
\ee

\section{Casimir-Polder interaction between two nanoparticles}
Let us now consider two nanoparticles, one located at the origin and one at 
$\mathbf{R}=(0,0,Z)$.  
Let the nanoparticles have both static electric and magnetic polarizabilities
$\bm{\alpha}_i$, $\bm{\beta}_i$, $i=1,2$.  We will again suppose the 
nanoparticles
to be anisotropic, but, for simplicity, having their principal axes aligned
with the direction connecting the two nanoparticles:
\be
\bm{\alpha}_i=\mbox{diag}(\alpha^i_\perp,\alpha^i_\perp,\alpha_z^i),\quad
\bm{\beta}_i=\mbox{diag}(\beta^i_\perp,\beta^i_\perp,\beta_z^i).
\ee
The methodology is very similar to that explained in the previous section.

\subsection{Electric polarizability}
\begin{figure}
\includegraphics{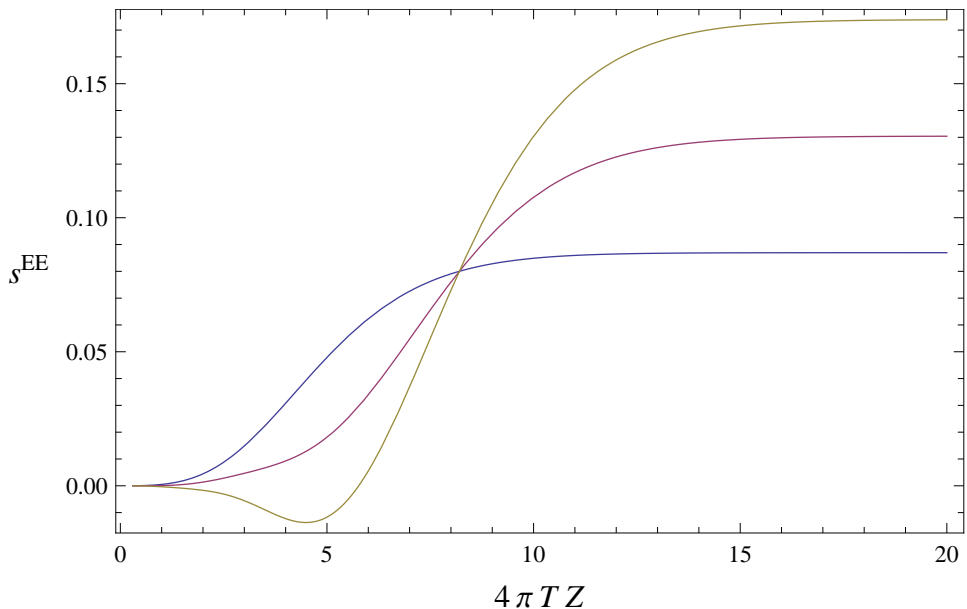}
\caption{\label{fig4} The entropy $s^{EE}(\gamma,y)$
for two anisotropic purely electrically
polarizable nanoparticles 
with separation $Z$ and temperature $T$.  When $\gamma=
\gamma_1\gamma_2>1$ the entropy can be negative.  The curves, bottom to top
for large $ZT$ are for $\gamma=0$ (blue), 1 (red), 2 (yellow), 
respectively. (Color online)}
\end{figure}

We start with the interaction between two electrically polarizable 
nanoparticles.  The free energy is
\be
F^{EE}=-\frac{T}2\sum_{m=-\infty}^\infty \tr [4\pi
\bm{\alpha}_1\cdot\bm{\Gamma}_0(
\mathbf{R})\cdot4\pi\bm{\alpha}_2\cdot\bm{\Gamma}_0(\mathbf{R})],\label{genfe}
\ee
where the free Green's dyadic is given in Eq.~(\ref{fgd}).  In view of
Eq.~(\ref{fgd2}), in terms of the polynomials (\ref{poly}), a simple
calculation yields ($y=4\pi Z T$)
\be
F^{EE}=-\frac{23}{4\pi Z^7}\alpha_z^1\alpha_z^2f(\gamma,y),
\ee
normalized to the zero-temperature Casimir-Polder energy
\cite{Casimir:1947hx}, where
\be
f(\gamma,y)=\frac{y}{23}\left[4\left(1-y\partial_y+\frac14y^2\partial^2_y
\right)+2\gamma\left(1-y\partial_y+\frac34y^2\partial_y^2
-\frac14y^3\partial_y^3+\frac1{16}y^4\partial_y^4\right)\right]\frac12
\coth\frac{y}2.
\ee
Here $\gamma=\gamma_1\gamma_2$, where $\gamma_i=\alpha_\perp^i/\alpha_z^i$.
The entropy is
\be
S^{EE}=\frac{23\alpha_z^1\alpha_z^2}{Z^6}s^{EE}(\gamma,y),
\quad s^{EE}(\gamma,y)=\frac\partial{\partial y}f(\gamma,y).\label{see}
\ee
The asymptotic limits are
\begin{subequations}
\bea
s^{EE}(\gamma,y)&\sim& \frac{2+\gamma}{23}, \quad y\gg1,\\
s^{EE}(\gamma,y)&\sim&\frac1{2070}(1-\gamma)y^3,\quad y\ll1,
\eea
\end{subequations}
so even in the pure electric case there is a region of negative entropy for
$\gamma>1$.  This is illustrated in Fig.~\ref{fig4}.
The coupling of two magnetic polarizabilities is given by precisely the same
formulas, except for the replacement $\bm{\alpha}\to\bm{\beta}$.

\subsection{EM cross term}
For the ``interference'' 
term between the magnetic polarization of one nanoparticle
 and the electric polarization of the other, we compute the free energy from 
the third term in Eq.~(\ref{ms}),
\be
F^{EM}=-\frac12\tr[\bm{\Phi}_0\cdot4\pi \bm{\alpha}_1\cdot
\bm{\Phi}_0\cdot4\pi\bm{\beta}_2]+(1
\leftrightarrow2).
\ee
This is easily worked out using the following simple form of the $\bm{\Phi}_0$
operator \cite{Milton:2011ed}:
\be
\bm{\Phi}_0(\mathbf{R})=-\frac{\zeta_m}{4\pi Z^3}\mathbf{R}\times(1+\zeta_m Z)
e^{-|\zeta_m|Z},\quad Z=|\mathbf{R}|.
\ee
The result for the free energy is
\be
F^{EM}=\frac{7}{4\pi Z^7}(\alpha_\perp^1\beta_\perp^2+\beta_\perp^1
\alpha_\perp^2)g(y),
\ee
which is normalized to the familiar zero temperature result \cite{feinberg},
where
\be
g(y)=\frac{y}{14}\left(y^2\partial_y^2-y^3\partial_y^3+\frac14y^4\partial_y^4
\right)\frac12\coth\frac{y}2.
\ee
The entropy is 
\be
S^{EM}=-\frac{7}{Z^6}(\alpha_\perp^1\beta_\perp^2+\beta_\perp^1
\alpha_\perp^2)s^{EM},\quad s^{EM}(y)=
\frac{\partial g(y)}{\partial y}.\label{sem}
\ee
This is always negative, vanishes exponentially fast for large $y$, and also
vanishes rapidly for small $y$,
\be
s^{EM}\sim -\frac{y^5}{7056}.
\ee
\subsection{General results}
We can present the total entropy for two nanoparticles having 
both electric and magnetic polarizabilities as follows,
\bea
S=\frac1{Z^6}\left[23\alpha^1_z\alpha^2_z
s^{EE}(\gamma^1_\alpha\gamma^2_\alpha,y)+
23\beta^1_z\beta^2_zs^{EE}(\gamma^1_\beta\gamma^2_\beta,y)
-7(\alpha_z^1\beta_z^2\gamma^1_\alpha\gamma^2_\beta+
\beta_z^1\alpha_z^2\gamma^1_\beta\gamma^2_\alpha)s^{EM}(y)\right],
\eea
where $s^{EE}$ and $s^{EM}$ are given by Eqs.~(\ref{see}) and (\ref{sem}), 
respectively.  For small $y$, the leading behavior of the entropy is
\bea
S&=&\frac{y^3}{90 R^6}[\alpha^1_z\alpha^2_z(1-\gamma^1_\alpha\gamma^2_\alpha)
+\beta^1_z\beta^2_z(1-\gamma^1_\beta\gamma^2_\beta)]\nn\\
&&\quad\mbox{}+\frac{y^5}{5040R^6}
[\alpha^1_z\alpha^2_z(4+7\gamma^1_\alpha\gamma^2_\alpha)
+\beta^1_z\beta^2_z(4+7\gamma^1_\beta\gamma^2_\beta)
+5(\alpha^1_z\beta^2_z\gamma^1_\alpha\gamma^2_\beta+
\beta^1_z\alpha^2_z\gamma^1_\beta\gamma^2_\alpha)+O(y^7).
\label{leadingterm}
\eea

\begin{figure}
\includegraphics{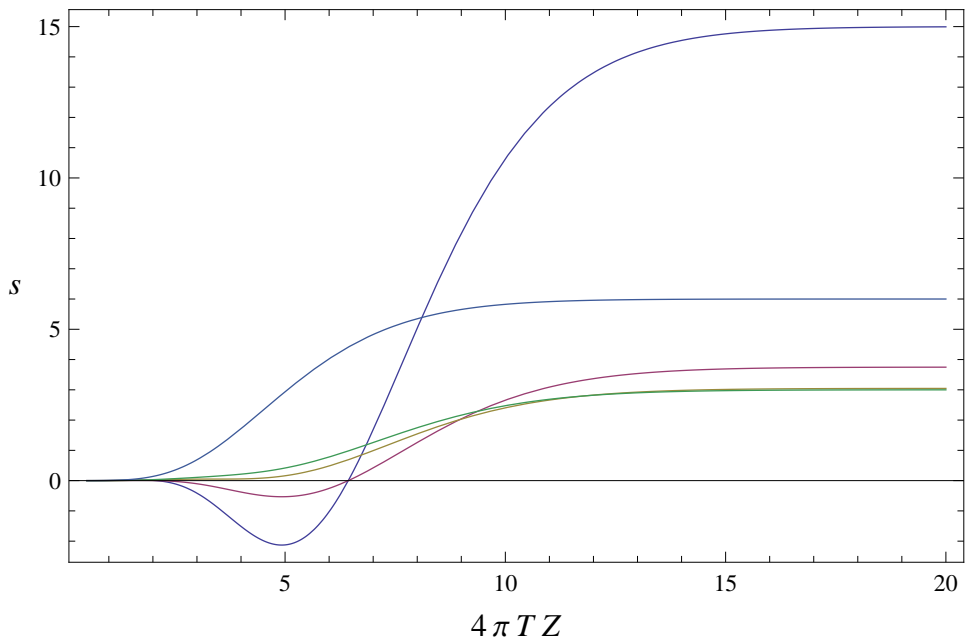}
\caption{\label{fig5} Entropy of two identical isotropic nanoparticles 
($\gamma_\alpha=
\gamma_\beta=1$) for different values of the ratio $r=\beta/\alpha$.  Starting
from highest to lowest curves on the left, the entropy is given for $r=1$ 
(purple), 0 (green),-1/8 (yellow), -1/2 (red), -2 (blue).  
What is plotted in this and the following figures is
$s$, where the entropy is $S=[(\alpha_z^1)^2/Z^6]s$. (Color online)}
\end{figure}

\begin{figure}
\includegraphics{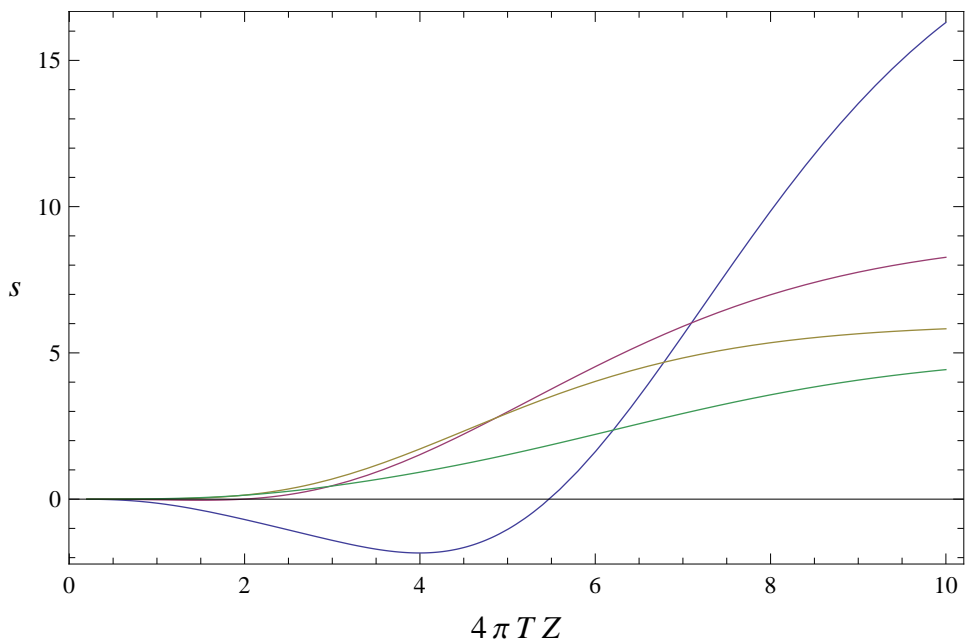}
\caption{\label{fig6} Here the 
identical nanoparticles have equal values of $\alpha_z=\beta_z$, 
and $\gamma_\beta=1$, but
different values of the electric anisotropy. Reading from bottom to top
on the right, we have $\gamma_\alpha=0$ (green), 1 (yellow), 2 (red), 4 (blue),
 respectively. (Color online)}
\end{figure}

\begin{figure}
\includegraphics{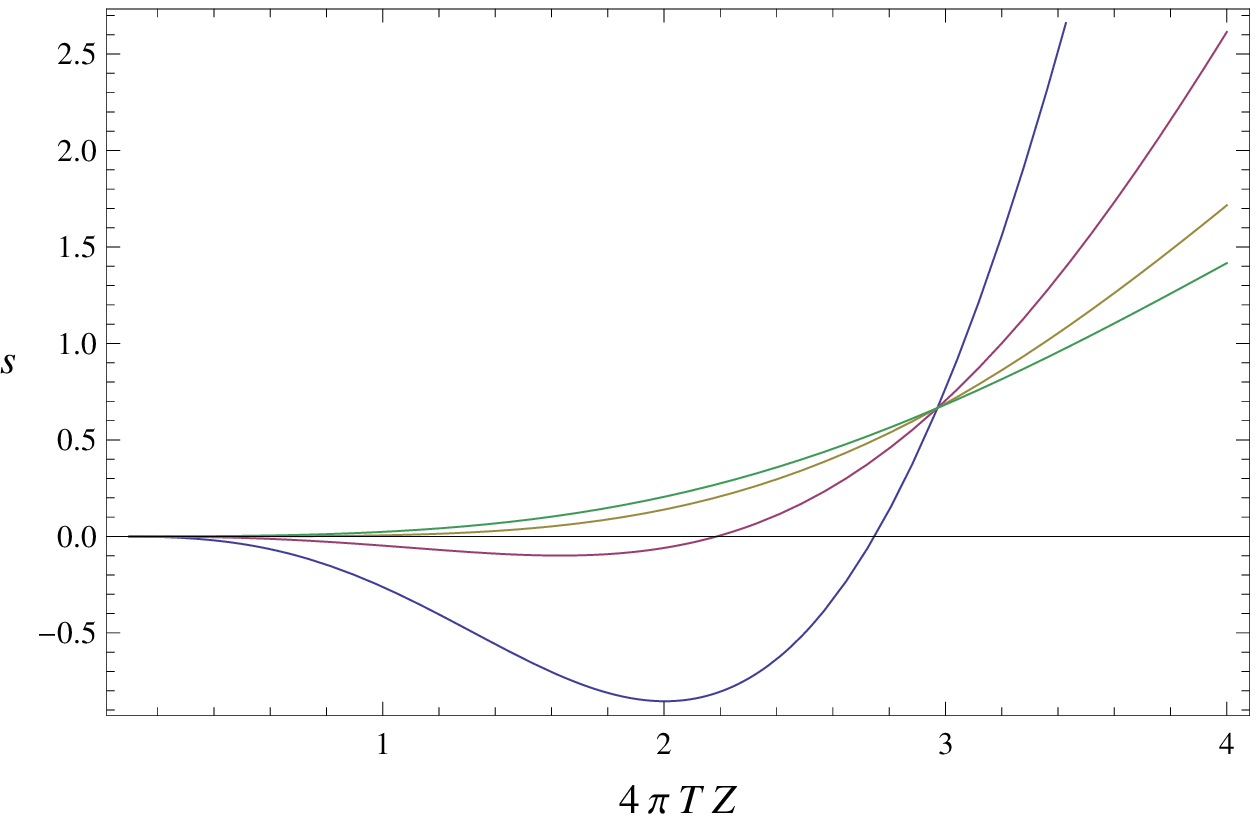}
\caption{\label{fig7} Here the identical nanoparticles
have equal electric and magnetic
polarizabilities, and equal anisotropies, which, starting from the bottom
on the right, have the values $\gamma=0$ (green), 1 (yellow), 2 (red), 
4 (blue), respectively. (Color online)}
\end{figure}

\begin{figure}
\includegraphics{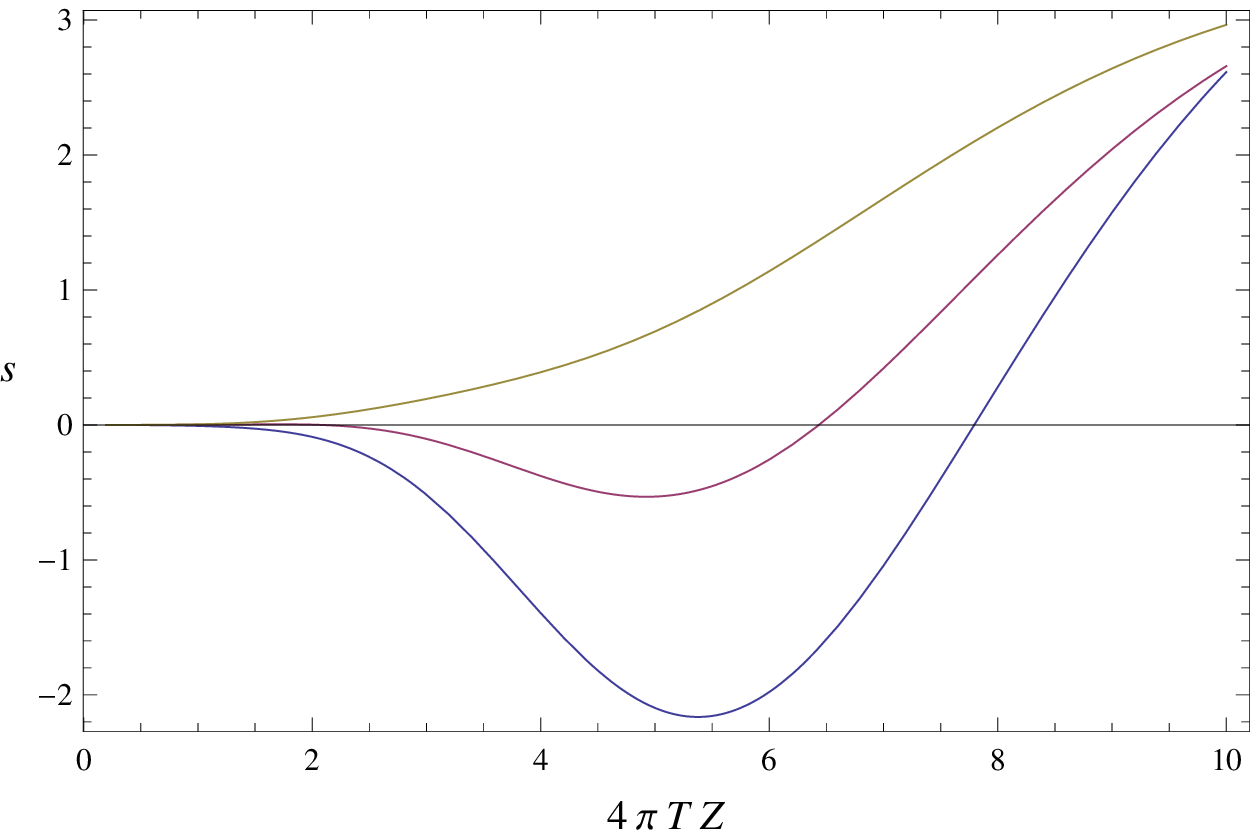}
\caption{\label{fig8} 
The case of two identical  conducting spheres where $\alpha_z=-2\beta_z$,
with electrical  isotropy, but magnetic anisotropy $\gamma_\beta=0$ (yellow), 
1 (red), 2 (blue), 
reading from top to bottom. (Color online)}
\end{figure}

\begin{figure}
\includegraphics{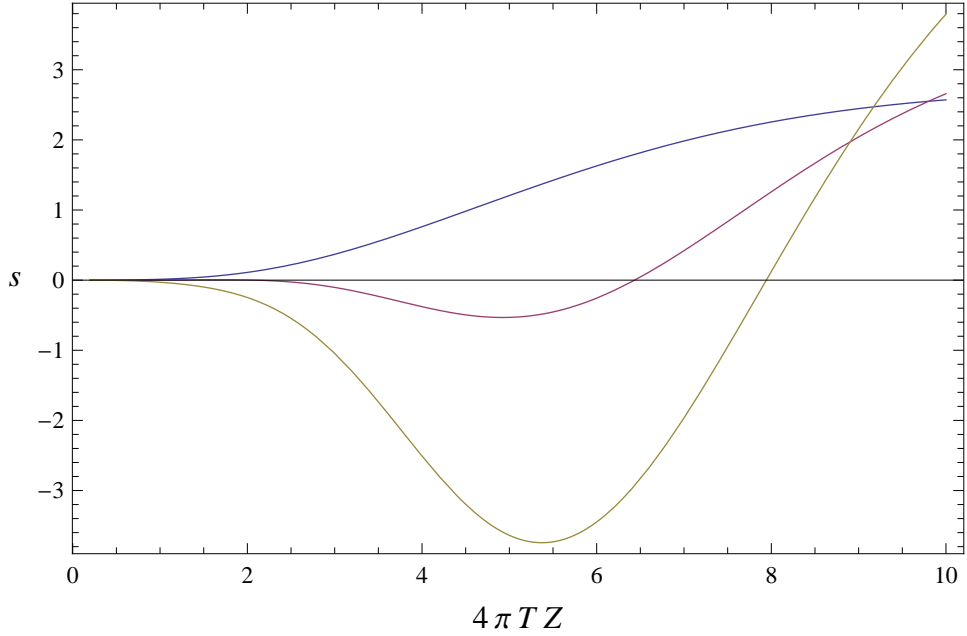}
\caption{\label{fig9}
The case of two identical conducting nanoparticles where $\alpha_z=-2\beta_z$,
with magnetic  isotropy, but electric anisotropy $\gamma_\alpha=0$ (blue), 
1 (red), 2 (yellow), 
reading from top to bottom in the middle. (Color online)}
\end{figure}

\begin{figure}
\includegraphics{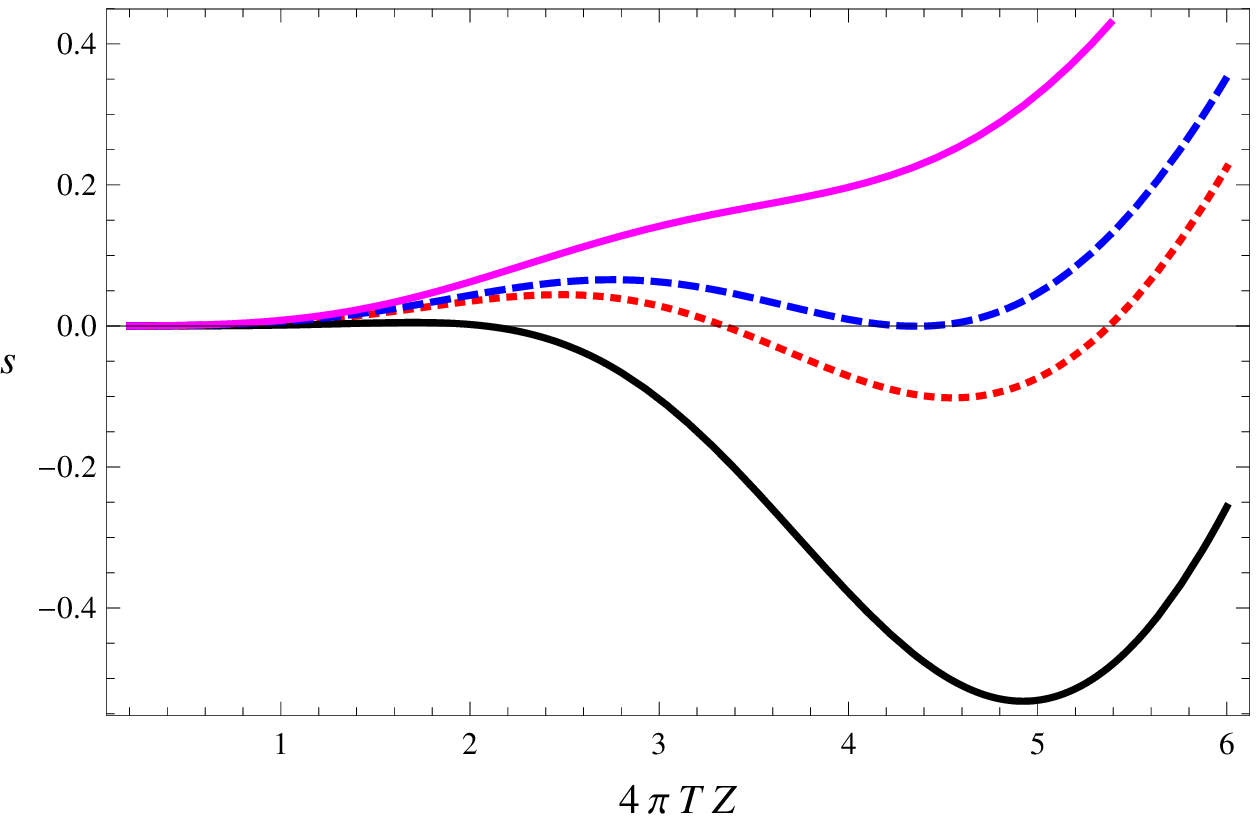}
\caption{\label{fig10} Two identical
nanoparticles  with $\beta_z=-\alpha_z/2$, appropriate for a 
conducting sphere, isotropic magnetically, but with electric anisotropies
$\gamma_\alpha= 0.6$ (magenta), 0.743 (dashed blue), 0.8 (short dashed red), 
1 (black), shown from top to bottom.}
\end{figure}

In the following six figures
 we present graphs of the entropy for the case of identical
nanoparticles, for simplicity, $\alpha_z^1=\alpha_z^2$, $\beta_z^1=\beta_z^2$,
$\gamma_\alpha^1=\gamma_\alpha^2$, $\gamma_\beta^1=\gamma_\beta^2$.
In Fig.~\ref{fig5} we show the entropy for isotropic nanoparticles 
with different
ratios of magnetic to electric polarizabilities; negative entropy appears
when the ratio is smaller than about $-1/8$.  This is a nonperturbative effect,
because the leading power of $y^3$ for small $y$ has a vanishing coefficient
in this case, and the $y^5$ term has a positive coefficient
---See Eq.~(\ref{leadingterm}).  (The radius of convergence
of the series expansions for the free energy is $|y|=2 \pi$.)  Thus,
perfectly conducting spheres, for which the ratio of magnetic to
electric polarizabilities is $-1/2$, exhibit $S<0$. 
In Fig.~\ref{fig6} we examine the case of equal $z$ components of the
 electric and magnetic 
polarizabilities, but when only the electric polarizability is anisotropic.
Negative entropy occurs when $\gamma_\alpha>1$, which we see perturbatively
from Eq.~(\ref{leadingterm}).
In Fig.~\ref{fig7} we consider the  nanoparticles 
as having equal polarizabilities
and equal anisotropies.  Again, as seen perturbatively, the boundary value
for negative entropy is $\gamma=1$.
The case of a conducting sphere has $\beta=-\alpha/2$.  We examine this 
situation in Fig.~\ref{fig8}, for different magnetic anisotropies, and in 
Fig.~\ref{fig9}, for different electric anisotropies.  In this case
the leading term in Eq.~(\ref{leadingterm}) vanishes at $\gamma=1$, 
so the appearance
of negative entropy for $\gamma\le1$ is nonperturbative.  
In fact, the boundary values for the two cases are $\gamma_\beta=0.5436$ 
and $\gamma_\alpha=0.7427$, respectively.  For the latter case, this is
illustrated in Fig.~\ref{fig10}.

\begin{figure}
\includegraphics{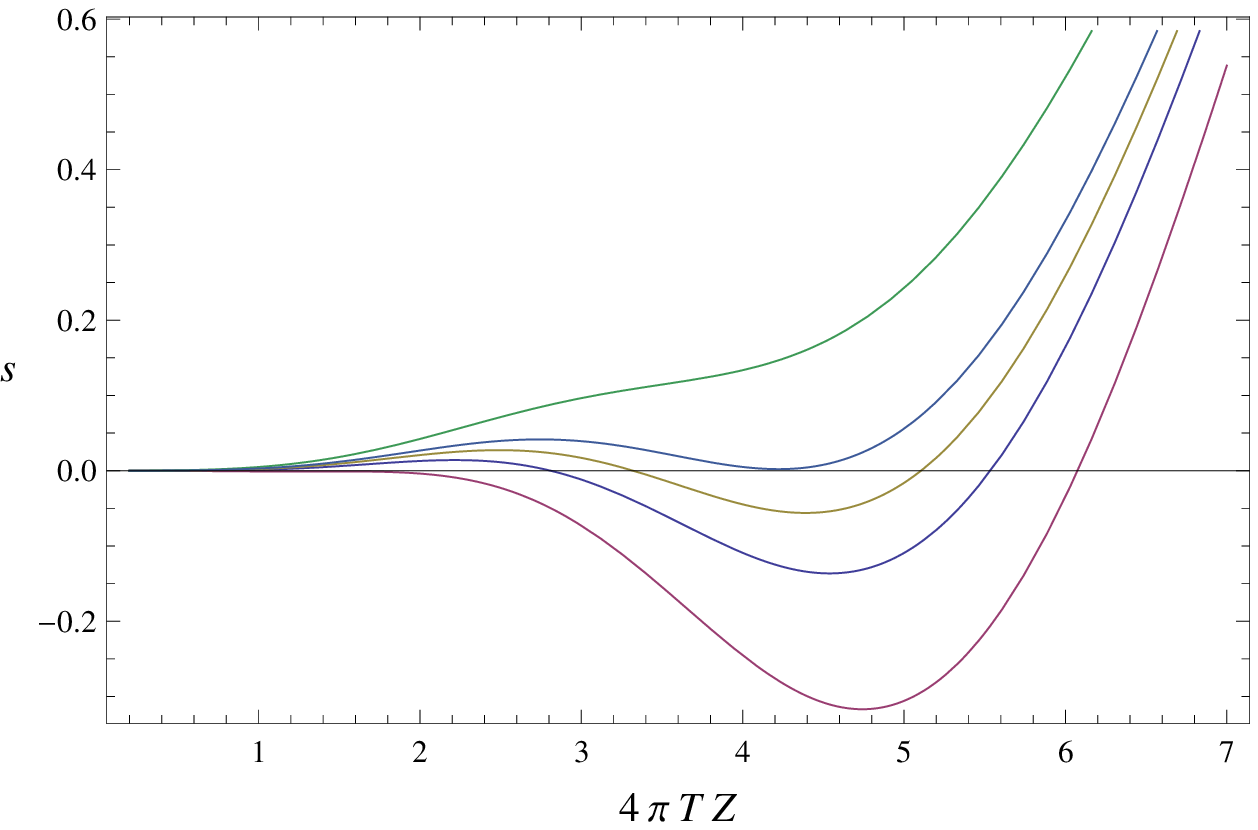}
\caption{\label{fig11} Interaction entropy between a small perfectly 
conducting nanoparticle, 
for which $\bm{\beta}_1=-\frac12\bm{\alpha}_1$, and a Drude nanoparticle 
with the same electric polarizability and no magnetic polarizability,
$\bm{\alpha}_2=\bm{\alpha}_1$, $\bm{\beta}_2=0$.  The electric
 anisotropies of the two nanoparticles 
are assumed equal, while the perfectly conducting nanoparticle is assumed
to have no magnetic anisotropy. The curves, top to bottom, are for anisotropy 
$\gamma_\alpha=0.8$ (green), $0.91$ (purple), $0.95$ (yellow), $1.0$ (blue), 
and $1.1$ (red), respectively. (Color online)}
\end{figure}

\begin{figure}
\includegraphics{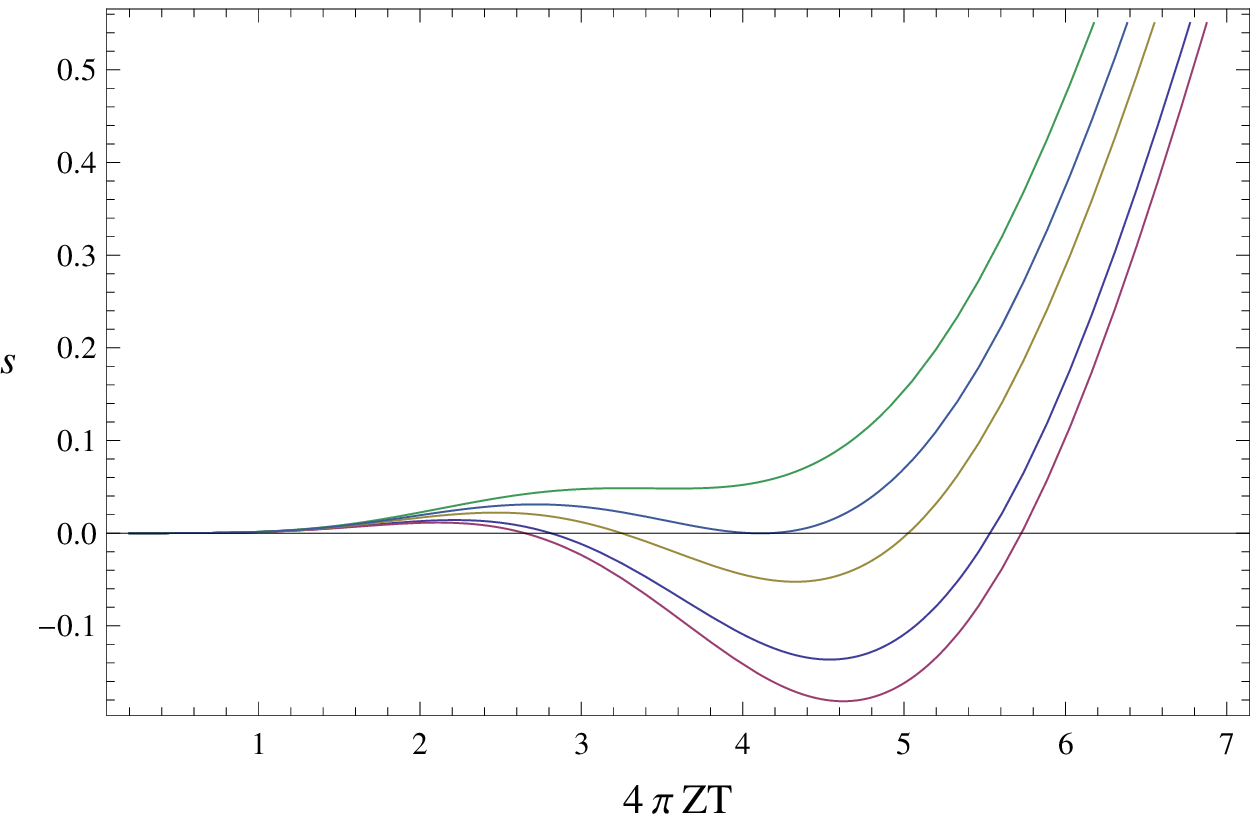}
\caption{\label{fig12} Interaction entropy between a small perfectly 
conducting nanoparticle, 
for which $\bm{\beta}_1=-\frac12\bm{\alpha}_1$, and a Drude nanoparticle 
with the same electric polarizability and no magnetic polarizability,
$\bm{\alpha}_2=\bm{\alpha}_1$, $\bm{\beta}_2=0$.  Now it is assumed that
the nanoparticles are electrically isotropic.  The dependence on the magnetic 
anisotropy of the first nanoparticle is shown.
Reading from top to bottom the magnetic anisotropies are
$\gamma_{\beta 1}=0.5$ (green), 0.66 (purple), 0.8 (yellow), 1 (blue), 
1.1 (red). (Color online)}
\end{figure}

An interesting case is the interaction of a perfectly conducting 
nanoparticle with a Drude nanoparticle, by which we mean that 
the latter  has vanishing magnetic
polarizability.  In Fig.~\ref{fig11} we consider the electric anisotropies
to be the same, while in Fig.~\ref{fig12} 
we show how the entropy changes as
we vary the anisotropy of the magnetic polarizability of the perfectly 
conducting sphere. For isotropic spheres there is always a region of negative
entropy.

\section{Conclusions}
In this paper we have studied purely geometrical aspects of the entropy that
arise from the Casimir-Polder interaction, 
either between a polarizable nanoparticle
and a conducting plate, or between two polarizable nanoparticles.  
 In all cases, the entropy vanishes at $T=0$, so the 
issues mentioned in the Introduction concerning the violation of the
Nernst heat theorem do not appear in the Casimir-Polder regime.
We consider the
simplified long distance regime where we may regard both the electric and
magnetic polarizabilities of the nanoparticles as constant in frequency.  
Thus, throughout we are assuming that the separations $Z$ are large compared
to the size of the nanoparticles, $a$.  This same restriction justifies
the use of the dipole approximation for the nanoparticles. It has
been known for some time that negative entropy can occur between a purely
electrically polarizable isotropic nanoparticle
 and a perfectly conducting plate.
Here we consider both electric and magnetic polarization for both the
nanoparticle
and the plate.  Negative entropy frequently arises, but requires interplay
between electric and magnetic polarizations, or anisotropy, in that the 
polarizability of the nanoparticles 
must be different in different directions.  
Interestingly,
although in some cases the negative entropy is already contained in the 
leading low-temperature expansion of the entropy, in other cases negative
entropy is a nonperturbative effect, not contained in the leading
behavior of the coefficients of the low temperature expansion.  
What we observe here extends what
has been found in calculations of the entropy between a
finite sphere and a plate.  We summarize our findings in Table \ref{tab},
which, we again emphasize, refer to the dipole approximation, appropriate
in the long-distance regime, $Z\gg a$.  Surprisingly, perhaps, negative
entropy is a nearly ubiquitous phenomenon:
Negative entropy typically occurs when a polarizable
nanoparticle is close to another such particle or to a conducting plate.
This is not a thermodynamic problem because we are considering only
the interaction entropy, not the total entropy of the system.  Nevertheless,
it is an intriguing effect, deserving deeper understanding.

 For confrontation with future experiments, the static 
approximation for the polarizabilites would have to be removed, a simple task 
in our general formalism. We are not aware if any present 
experiments concerning Casimir energies between
nanoparticles and surfaces, but we hope this investigation will spur efforts
in that direction.
\begin{table}
\centering
\begin{tabular}{lr}
Nanoparticle/nanoparticle \\
or nanoparticle/plate& Negative entropy?\\[0.5ex]
\hrulefill&\hrulefill\\
E/E& $S<0$ occurs for $\gamma_\alpha>1$\\
E/M &$S<0$ always\\
PC/PC&$S<0$ for $\gamma_\alpha>0.74$ or $\gamma_\beta>0.54$\\
PC/D&$S<0$ for $\gamma_\alpha>0.91$ or $\gamma_\beta>0.66$ \\
E/TE plate&$S<0$ always\\
E/TM plate&$S<0$ for $\gamma_\alpha>2$\\
E/PC or D plate&$S<0$ for $\gamma_\alpha >1/2$\\
\hrulefill&\hrulefill\\

\end{tabular}
\caption{\label{tab} The table shows when a negative entropy region can occur, 
in different situations.
Here E refers to an electrically polarizable particle, M a magnetically
polarizable particle, PC means a perfectly conducting particle or plate, 
D means an object described by the Drude model.  TE and TM refer to the
transverse electric and transverse magnetic contributions to a perfectly
conducting plate.  The electric (magnetic) anisotropy is defined by 
$\gamma_\alpha=\alpha_\perp/\alpha_z$ ($\gamma_\beta=\beta_\perp/\beta_z$).
Analogous results can be obtained for other cases by electromagnetic duality.}
\end{table}

\acknowledgments
KAM and G-LI thank the Laboratoire Kastler Brossel for their hospitality
during the period of this work.  CNRS and ENS are thanked for their support.
KAM's work was further supported in part by  grants from the
 Simons Foundation and the Julian Schwinger Foundation.

\end{document}